\documentclass[review]{elsarticle}

\usepackage{amsmath}
\usepackage{algpseudocode}
\usepackage{enumitem}
\usepackage{float}
\usepackage{graphicx}
\usepackage{epsfig}
\usepackage{epstopdf}
\usepackage{tabularx}
\usepackage{color}
\usepackage{subfig}
\usepackage{comment}
\usepackage[normalem]{ulem}
\usepackage{setspace}
\usepackage{listings}
\usepackage{xcolor}
\usepackage{algorithm}
\DeclareGraphicsExtensions{.eps}
\usepackage{hyperref}

\newcommand{\squishlist}{
  \begin{list}{$\bullet$}
  { \setlength{\itemsep}{0pt}      \setlength{\parsep}{-0pt}
    \setlength{\topsep}{4pt}       \setlength{\partopsep}{0pt}
    \setlength{\listparindent}{-2pt}
    \setlength{\itemindent}{-5pt}
    \setlength{\leftmargin}{1em} \setlength{\labelwidth}{0em}
    \setlength{\labelsep}{0.5em} } }

\newcommand{\squishend}{
    \end{list}  }

\newcommand{\cred}{\textcolor{red}}

\makeatletter

\definecolor{mGreen}{rgb}{0,0.6,0}
\definecolor{mGray}{rgb}{0.5,0.5,0.5}
\definecolor{mPurple}{rgb}{0.58,0,0.82}
\definecolor{backgroundColour}{rgb}{0.95,0.95,0.92}

\lstdefinestyle{SQL} 
{ 
   language=sql, 
   framexleftmargin=5mm, 
   frame=shadowbox, 
   numbers=left, 
   basicstyle=\ttfamily, 
   numberstyle=\ttfamily\tiny, 
   numbersep=5pt, 
   tabsize=4, 
   breaklines=true 
}

\lstdefinestyle{CStyle}{
	commentstyle=\color{mGreen},
	keywordstyle=\color{magenta},
	stringstyle=\color{mPurple},
	basicstyle=\footnotesize,
	breakatwhitespace=false,         
	breaklines=false,
    basicstyle=\ttfamily\small, 
    frame=tblr,
	captionpos=bc,                    
	keepspaces=true,                 
	showspaces=false,                
	showstringspaces=false,
	showtabs=false,                  
	tabsize=4,
	caption=Communication message type,
	label=list1,
	language=C
}

\begin{document}

\begin{frontmatter}

\title{FT-LADS: Fault-Tolerant Object-Logging based Big Data Transfer System using Layout-Aware Data Scheduling}


\author[ajou]{Preethika Kasu}
\author[sogang]{Taeuk Kim}

\author[sogang]{Youngjae Kim\corref{mycorrespondingauthor}}
\author[kisti]{Jung-Ho Um}
\author[kisti]{Kyongseok Park}
\author[oak]{Scott Atchley}

\address[ajou]{Ajou University, Suwon Korea} 
\address[sogang]{Sogang University, Seoul Korea} 
\address[kisti]{Korea Institute of Science and Technology Information, Daejeon Korea}
\address[oak]{Oak Ridge National Laboratory, TN USA}

\address{\small{{kasu@ajou.ac.kr, \{taeugi323, youkim\}@sogang.ac.kr}, \{jhum, gspark\}@kisti.re.kr}, atchleyes@ornl.gov}

\cortext[mycorrespondingauthor]{Corresponding author\\
Author(s) declare(s) that there is no conflict of interest regarding the publication of this paper.}

\begin{abstract}
Layout-Aware Data Scheduler (LADS) data transfer tool, identifies and addresses the issues that lead to congestion on the path of an end-to-end data transfer in the terabit network environments. It exploits the underlying storage layout at each endpoint to maximize throughput without negatively impacting the performance of shared storage resources for other users. LADS can avoid congested storage elements within the shared storage resource, improving input/output bandwidth, and hence the data transfer rates across the high speed networks. However, absence of FT (fault tolerance) support in LADS results in data retransmission overhead along with the possible integrity issues upon errors. In this paper, we propose object based logging methods 
to avoid transmitting the objects which are successfully written to Parallel File System (PFS)
at the sink end. Depending on the number of logger files created, for the whole dataset, we classified our fault tolerance mechanisms into three different categories: {\em File logger}, {\em Transaction logger} and {\em Universal logger}. Also, to address space overhead of these object based logging mechanisms, we have proposed different methods of populating logger files with the information of the completed objects. We have evaluated the data transfer performance and recovery time overhead of the proposed object based logging fault tolerant mechanisms on LADS data transfer tool. Our experimental results show that,  LADS in conjunction with proposed object based fault tolerance mechanisms exhibit an overhead of less than $1\%$ with respect to data transfer time and total recovery time overhead is around $10\%$ of total data transfer time at any fault point. 
\end{abstract}


\end{frontmatter}

\section{Introduction}
\label{sec:intro}

Datasets grow rapidly - majorly due to the large-scale scientific simulations~\cite{lads17, raghul:fast14, sc14:sop} and also due to the growth of data capable Internet of Things (IoT) devices such as mobile devices, software logs, cameras, microphones, and wireless sensor networks~\cite{big_wiki}. The world's technological per-capita capacity to store information has roughly doubled every 40 months since the 1980s~\cite{science-info-cap-published}.
By 2025, International Data Corporation (IDC) predicts there will be 163 Zettabytes of data. While the the sheer size of the data, of course, is a major challenge, there exist other challenges to be answered in order to fully realize the potential benefits of big data. They include storage I/O bottleneck and high data movement cost between advanced computational centers.

To support an increase in the data volume, data centers are equipped with adequate storage capacity. But at times, it is necessary to access additional resources located at geographically distributed data centers. This requires, transferring huge volumes of data between data centers. Also, to ensure the availability of data in realtime, it is necessary to transfer the data at high transfer rates. Data transfer between data centers involves high speed networks as well as physical storage media such as hard disk drives. 
With the growing networking hardware capabilities, it is possible to achieve higher data transfer rates. But this is not sufficient for achieving higher end-to-end data transfer rate, which includes slower storage infrastructure. 

Even though networking hardware capabilities reach terabit speeds and storage capabilities reach exabytes, there is a clear mismatch between network to storage speeds. This poses a major challenge in achieving higher end-to-end data transfer rates. In order to reduce the impedance mismatch between network and storage and to improve the scalability, distributed file system, parallel file systems (PFS) is used. PFS uses different servers to service metadata and I/O operations in parallel. Also, in order to improve the throughput, PFS uses higher number of I/O servers connected with more disks. Typically, large-scale storage systems use tens to hundreds of I/O servers, each with tens to hundreds of disks. 

Though, PFS significantly improves scalability and performance, due to the fact that, these storage systems are shared resources between multiple clients, it is possible to content for the same resource by multiple clients. As contention for resources increases, there can be serious gap between expected and observed I/O performance by users~\cite{Gulati:2007:PAC:1254882.1254885, Xie:2012:COB:2388996.2389007}. Also, it is possible that, at times, some of the disks are overloaded while most or not. This kind of load imbalance is quite a serious problem in parallel file systems~\cite{Welch:2008:SPP:1364813.1364815}. With these observations, researchers have proposed a new bulk data transfer framework called, Layout Aware Data Scheduler (LADS), which avoids temporarily congested servers during data transfers~\cite{kim:fast15}~\cite{lads17}. LADS implemented using CCI for communication~\cite{atchley11:cci, cci-forum}.
LADS exploits the underlying storage layout at source and sink to maximize throughput without negatively impacting the performance of shared storage resources for other users.
LADS focuses on objects, rather than files, which allows the LADS framework to implement layout-aware scheduling algorithms. 
Due to 
this object level scheduling, objects may be transferred out-of-order from source to sink. 

One of the major challenges in distributed environments is failure; hardware, network, and software might fail at any point of time. And it is very costly to retransmit the whole data from the beginning while transmitting several terabytes of data. In addition, fault in the middle of data transfer may cause transaction failure before the entire file has been transferred.
In particular, it is very costly to send data back from the beginning when the system fails while transmitting several terabytes of data. In addition, a system failure in the middle of a data transfer may cause a partial transaction failure before the entire file has been transferred.
If any object in LADS is lost due to any of the faults along the end-to-end path, this will result in data integrity and performance issues. 
However,
the current LADS implementation does not offer any solution to the faults occurred in the end-to-end path. The absence of fault tolerance mechanisms~\cite{ft-impl, ft-checkpoint} resulting in retransmitting the whole file (or entire objects which have been transferred so far) upon fault, causing unnecessary congestion~\cite{ft-sw, ft-book}. 


Due to \textit{out-of-order} nature of object transmission, checkpoint based logging file offset (or) logging the index of last object being transferred is not enough for resuming the transfers upon fault \cite{bbcp}.
Another approach is to maintain the log of all objects that were successfully sent and written at the sink end PFS. However, this kind of logging mechanism~\cite{logging} will have an impact on the overall space occupied by the logger and also the amount of time consumed to log the object information while transferring the data and to retrieve the successfully completed object information upon fault. These factors will have direct impact on the overall performance of the data transfer. Our main objective is to design object based fault tolerance mechanism to minimize the time, space and retrieval overhead while not negatively impacting the performance of data transfer.

In this paper, 
we propose
{\em object based file logging fault tolerance mechanism}, to use in conjunction with LADS. 
In order to analyze the performance and space overhead of the fault tolerance mechanisms on LADS, 
we propose different object based fault tolerance logging mechanisms. This paper makes the following contributions.
\begin{itemize}
	\item
	In object based logging, each and every file in dataset is associated with one logger file. Growing the size of the dataset, the number of logger files will also increase. Increase in the number of logger files causes non-negligible overhead to file system. On creating a file, system-wide open file table in kernel needs to be updated, and per-process file table is also updated. If threads concurrently request to update the shared table, it will cause contention. To avoid this, light-weight logging mechanism is implemented.
	
	\item
	Depending on the number of logger files generated per dataset, we propose three different object based fault tolerance mechanisms: 
	\textit{File Logger}, \textit{Transaction Logger}, and \textit{Universal Logger}~\cite{ft-poster}. 
	In case of file logger mechanism, each file in the target dataset is associated with one logger file, which will be used for recovery upon resuming from the fault. Whereas, in case of transaction and universal logger mechanisms, one logger file is associated with one transaction and whole dataset respectively.
	
	\item
	
	The space overhead varies depending on how the completed objects information is populated in the logger files. We propose different logging methods: \textit{char}, \textit{int}, \textit{enc}, \textit{binary}, \textit{bit8} and \textit{bit64}~\cite{ft-poster}. All these methods are evaluated for space overhead with the above mentioned logger mechanisms.
	\item
	We have analyzed performance overhead of object based fault tolerance mechanism(s)/method(s) with respect to performance and space overhead. 
	For evaluating our implementation, we have used Lustre filesystem based nodes which communicate over InfiniBand (IB) network. From our evaluation results, we have observed space overhead of around 60~KB (KiloBytes), data transfer time
	overhead of less than $1\%$ and total recovery time overhead is around $10\%$ of total data transfer time at any fault point.
	
\end{itemize}

The rest of the paper is organized as the following: Section~\ref{sec:back} describes LADS background followed by the motivation of our work. Section~\ref{sec:ladsarch} reviews LADS system implementation details. Section~\ref{sec:logger} presents the proposed object based logging mechanisms to support fault tolerance with LADS. Section~\ref{sec:Fault} describes the design and architectural changes incorporated in LADS to support fault tolerance. The experimental results and related works are presented in Section~\ref{sec:expr} and Section~\ref{sec:related}. We conclude the paper in Section~\ref{sec:conc}.

\section{Background and Motivation}
\label{sec:back}

\subsection{Layout-aware Data Scheduling}

High speed networks and slow storage servers are involved while transferring the data between data centers~\cite{kim:fast15}~\cite{lads17}. Storage server might experience congestion if the number of I/O requests exceed storage server capability. Due to this congestion, storage server consumes more time to service new I/O request. This kind of behaviour is common and is expected with parallel file system (PFS) when multiple applications (or) single large application is trying to access files on the same OST\footnote{We use Lustre terminology for object storage servers (OSS) and targets (OST). An OST manages a single device. A single Lustre OSS manages one or more OSTs.}. To some extent, it is possible to avoid this kind of congestion issues by using OS caching and application level buffering techniques. But big data transfer tools~\cite{bbcp}~\cite{sc12:rftp} can not benefit from this kind of techniques, due to the high volume of the data. While transferring the data, if the source end of the transfer is congested, due to large read requests, source end of the application will not be able to feed the data to network buffers at the expected rate, causing network buffers to drain and hence stall the transfer. On the other hand, if the sink end is congested, due to large write operations, sink end will not be able to consume the data at the expected rate, causing its buffers to full and eventually stall the I/O threads at source end due to unprocessed buffers.

Existing big data transfer tools~\cite{bbcp}~\cite{ Allcock:2005:GSG:1105760.1105819} consider the workload in-terms of logical files, they do not consider how the file is physically distributed. If single I/O thread is assigned to transfer the file, it will work on the file sequentially till the whole file is read or write. This way it will take long time to transfer all the files in the dataset, as one file is transferred at a time. To improve the data transfer performance, it is possible to assign multiple I/O threads to process the data transfer. Employing multiple I/O threads without the knowledge of the physical distribution of the file, might result in disk contention issues as multiple threads compete for the same OSS or OST. Due to this contention, data transfer performance of the application will be degraded. LADS~\cite{kim:fast15}~\cite{lads17} data transfer tool addresses storage contention issues by considering the physical distribution of the file over different OSTs. LADS considers the workload as objects rather than files. Hence, workload is divided into $O$ objects, where $O$ is the objects of $N$ total files, and each object represents one transfer maximum transmission unit ($MTU$) of data. LADS avoids the OST contention by scheduling accesses of OSTs. Because of this, objects of any file can be transferred before objects of another file.

In parallel file system (PFS), file is stripped over multiple OSTs to improve the overall I/O throughput. LADS improves the data transfer performance by exploiting the PFS layout. As the file is distributed over N OSTs, LADS employs N threads to request N objects each from separate OST. If any of the request is delayed by a congested OST, the N-1 threads are free to issue new requests to other OSTs. By the time, request to the slow server completes, other threads of LADS might be able to retrieve more than N objects. With this, the overall data transfer parallelism and hence the data throughput would be improved. Though, LADS tool exhibits higher throughput than existing tools, due to the lack of fault tolerance support to handle software, hardware or network failures during the transfer, would need to retransmit all the objects of whole dataset upon recovery from failure.

\subsection{Motivation}
\begin{figure}[!t]
	\centering
	\begin{tabular}{@{}cccccccccccccc@{}c@{}}
		\includegraphics[width=0.38\textwidth]{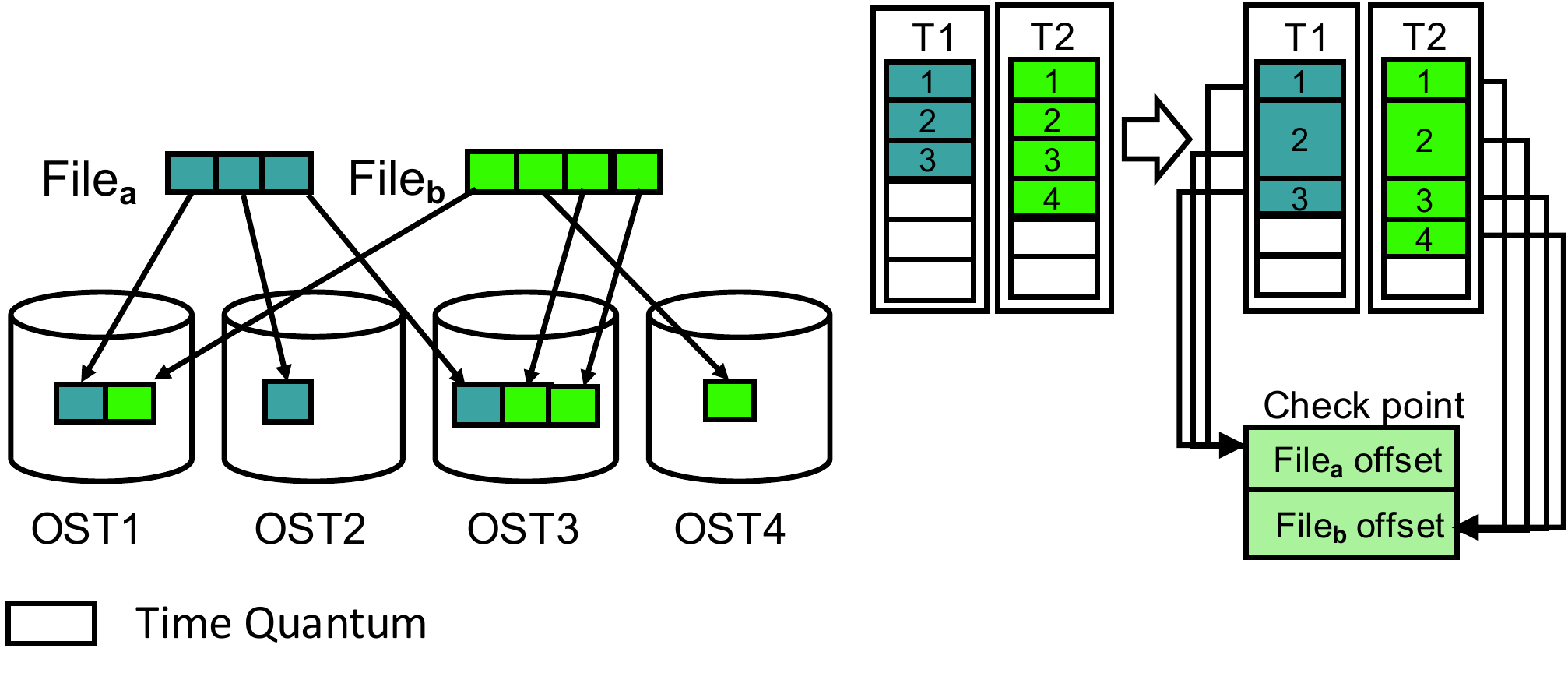} &
		\hspace{0.2in}		
		\includegraphics[width=0.4\textwidth]{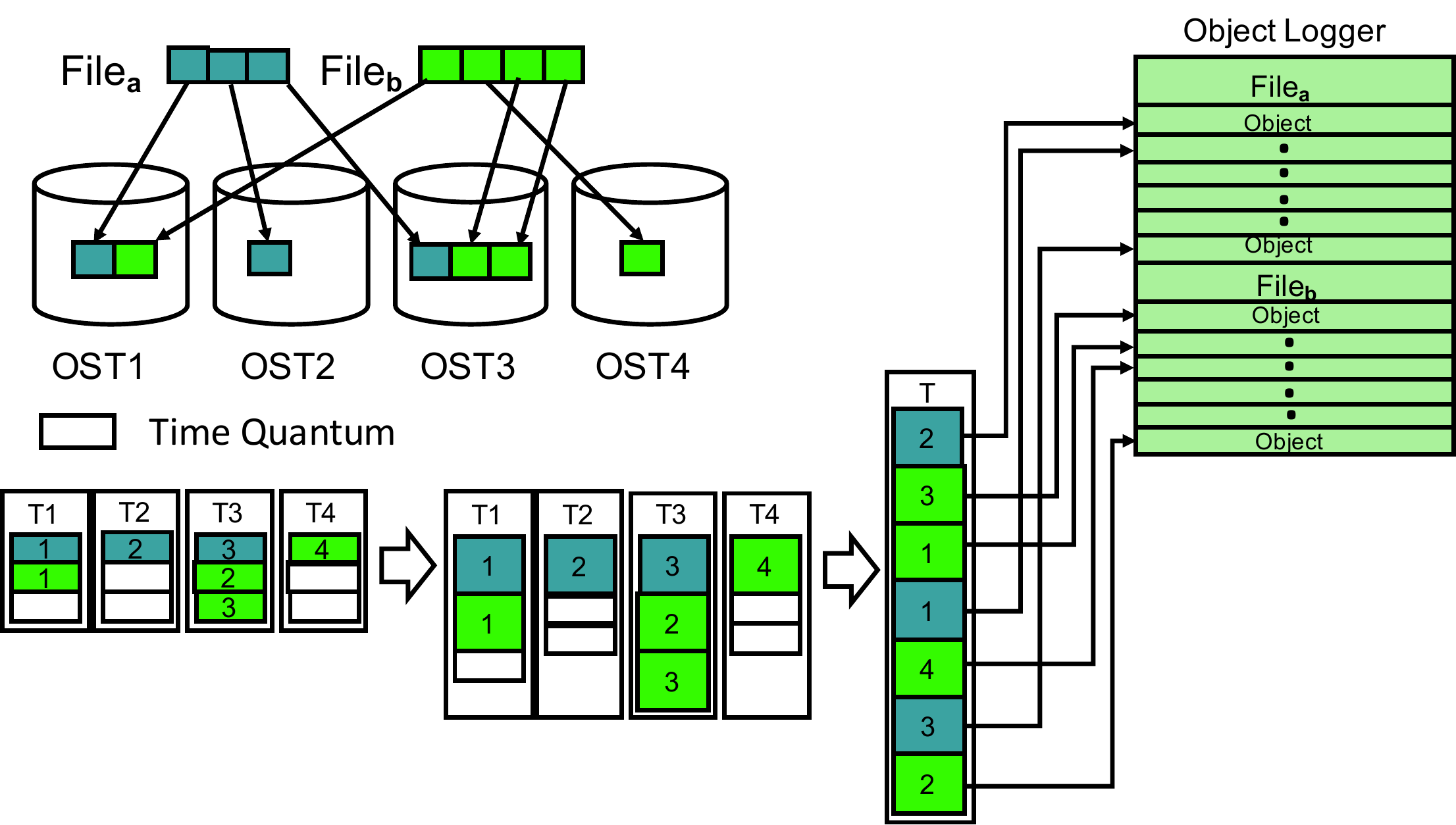} \\		
		\small (a) File based offset logging  &
		\small (b) Object based logging  \\		
	\end{tabular}
	\caption{\small File based and Object based Logging}
	\label{fig:logging_back}
\end{figure}

Traditional big data transfer tools, like bbcp, rely on logical view of the files, which ignores the underlying system architecture. Due to this, objects of the same file are transferred in sequence. As shown in Figure~\ref{fig:logging_back}(a), even though there is a possibility of resource contention between threads, T$_1$ and T$_2$, all the objects of File$_a$ and File$_b$ are transferred in sequence. Thread T$_1$ transfers the first object of the File$_a$ and then records file offset information. After completing the second object, overwrite the checkpoint record with the updated file offset information.

This will continue for all files in the dataset. During this process, if there is any fault, transfer tool will check for checkpoint record and if it exists for the target file, start transferring the objects beginning from the offset found in the checkpoint record. 	

In contrast to traditional data transfer tools, to avoid unwanted OST resource contentions, LADS exploit the underlying storage architecture and view the files in physical point of view. 
LADS consider the entire workload of $O$ objects, where $O$ is all of the objects in the $N$ total files, and each object represents one transfer MTU of data.
As shown in Figure~\ref{fig:logging_back}(b), a thread can be assigned to an object of any file on any OST without requiring all objects of a particular file be transferred before objects of another file. 
From the Figure~\ref{fig:logging_back}(b), we can observe that the second object of File$_a$ is transferred first and then the first object is transferred. Similarly, we can observe the out-of-order object transfer for File$_b$ too. Similar kind of mechanism will be continued for all files in the dataset. As objects are transferred out of order, this is not possible to recover the completed object information by logging checkpoint based file offset as shown in Figure~\ref{fig:logging_back}(a). 
Hence, we need to device a mechanism by which we can retrieve all the completed objects that are successfully transferred prior to the fault. To achieve this, one method is to maintain the information of all objects of all logical files, that are successfully transferred.

But, here the major issue is the amount of space occupied by the log files in case of big dataset and also the logging overhead on the data transfer rate.
So 
our work is motivated to answer the following questions:

\begin{itemize}
	\item{How to minimize the object based FT overhead on LADS data transfer rate?}
	\item{How to minimize the number of files created while processing data transfer?}
	\item{How to minimize the space occupied by the object based FT method?}
	\item{How to reduce the recovery time while resuming the transfer?}
\end{itemize}

To address the above challenges, we have proposed object based fault tolerance mechanisms to be used in conjunction with LADS tool.

\section{LADS Architecture}
\label{sec:ladsarch}

In this section, we first describe LADS system implementation details. Next, we discuss the possible performance issues with LADS tool in faulty environments.

\subsection{LADS Overview}
LADS~\cite{kim:fast15}~\cite{lads17} system is implemented by having one {\it{master}} thread, configurable number of {\it{I/O}} threads and one {\it{comm}} thread. 
{The \it{master}} thread is responsible for scheduling the objects transfer, whereas {\it{I/O}} threads read or write the object data from or to PFS.
{The \it{comm}} thread handles the communication between source and sink. 
{The \it{master}} and {\it{I/O}} threads block while waiting for a resource, however, {\it{comm}} thread always progresses the communication between source and sink.

Upon initiating the transfer, source and sink processes (hereafter simply source and sink) initialize the threads necessary for the communication along with all the required locks, wait queues, OST work queues and allocate RMA buffers used for data transfer. The {\it{comm}} thread, which communicates using CCI, opens a CCI endpoint and registers RMA buffer with CCI. Sink end {\it{comm}} thread opens and waits for the connection from the source. The source {\it{comm}} thread establishes connection with the open sink end CCI endpoint. During connect request, source {\it{comm}} thread, sends its maximum object size, number of objects in the RMA buffer, and the memory handle for the RMA buffer. The sink {\it{comm}} thread accepts the connection request, which triggers the CCI connect event on the source. 

\begin{figure}[!t]
	\begin{center}
		\includegraphics[width=0.95\textwidth]{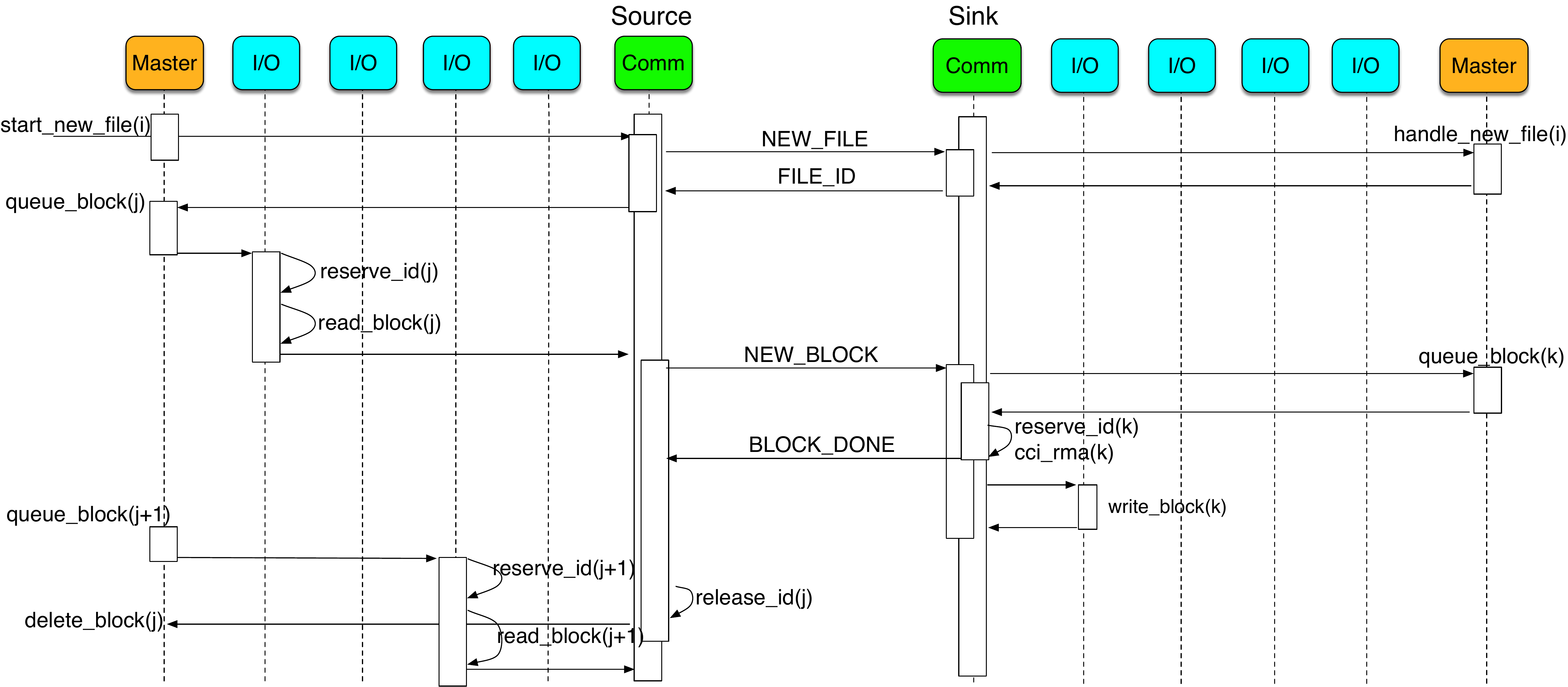}
		\caption{LADS data transfer sequence diagram }
		\label{fig:lads_seq}
	\end{center}	
\end{figure}

Sequence flow of data transfer between source and sink endpoints is shown in Figure~\ref{fig:lads_seq}. For each file in the target dataset, the source {\it{master}} thread generates NEW\_FILE request and enqueues the same with the work queue of the {\it{comm}} thread. {Source \it{comm}} thread dequeues the request and transfers the same to sink end using CCI interface. At sink end, {\it{comm}} thread receives NEW\_FILE request and enqueues the same to {\it{master}} thread's work queue and wakes it up. Based on the target file information in the request, {\it{master}} thread opens the file and adds the file descriptor to the FILE\_ID request and then enqueues the same on {\it{comm}} thread's work queue. {Sink end \it{comm}} thread dequeues the request and sends it to source. On receiving the FILE\_ID request, {\it{comm}} thread enqueues the request on {\it{master}} thread's wait queue and wakes it up. The source {\it{master}} thread splits the file as per object size and generates NEW\_BLOCK request and enqueues the request on {\it{I/O}} thread wait queue and wakes it up. 
Based on the first NEW\_BLOCK request, {\it{I/O}} thread determines the OST to be used for reading the object data and issues pread() to read the object data into the RMA buffer registered with CCI. 
On completing the read operation, it enqueues the request on the {\it{comm}} thread's work queue. The {\it{comm}} thread dequeues the request and transfers the same to sink. 

At sink end, {\it{comm}} thread receives the NEW\_BLOCK request and then tries to reserve the RMA buffer. If RMA buffer is available, then initiates RMA read operation. If it fails to get RMA buffer, it enqueues the request on {\it{master}} thread's work queue. {The \it{master}} thread waits till RMA buffer is available. Once the buffer is available, {\it{master}} thread enqueues the request on {\it{comm}} thread's queue, which  issues RMA read operation. Upon successful RMA read, sink {\it{comm}} thread sends BLOCK\_DONE request to source and wake up an {\it{I/O}} thread. The {\it{I/O}} thread dequeues the request, calls pwrite() to write the data to disk. On completing write operation, {\it{I/O}} thread releases the RMA buffer, so the {\it{comm}} thread can initiate another RMA Read. This process is repeated till all the objects of the dataset are successfully transferred to sink.

\subsection{Problem Definition}
{\em Fault Tolerance:} Often, during large transfers, the connection between the transferring systems is lost. Connection errors might be the result of software, hardware or network failures. The data transfer tool ability to resume the transfer from where it let off avoids transferring already completed files (or) objects before connection failure. This not only avoids congestion due to redundant file transfer but also improves the overall transfer performance in faulty network conditions. LADS object transfer protocol, considers whether the object is successfully read (sink) from RMA buffer or not. If there is any error while writing to PFS, it will go unnoticed and hence the transferred data will not be useful for further analysis due to data corruption. Also, if there is any error during the transfer, LADS object transfer protocol restarts the transfer from the beginning. This will not only waste the resources but also increase the overall transfer time. We propose solutions to avoid the transfer redundancy by implementing light-weight object based logging fault tolerance mechanisms with LADS tool.

\section{Object based Logging}
\label{sec:logger}
As LADS employs Layout-aware and OST congestion aware I/O scheduling algorithms, objects of any file on any OST can be transferred before objects of another file. Because of this, it is possible that, objects of same logical file might be transferred out of order. Due to this \textit{out-of-order} nature, logging file offset based fault tolerance mechanisms can not be employed with LADS. To support fault tolerance with LADS, information of all the objects of a logical file that are successfully written to PFS at sink need to be maintained. This process can not only computationally expensive but also result in additional space overhead.
In order to minimize the computational and space overhead, we have proposed different logging mechanisms.

In this section, we describe the proposed object based logging mechanisms~\cite{ft-poster} to support fault tolerance with LADS. 

\subsection{Object based logging Mechanisms}
Depending on the number of logger files generated per dataset, 
we propose three different object based fault tolerance mechanisms: 
File Logger, 
Transaction Logger, 
Universal Logger.

\subsubsection{File Logger}
Using the file logger object based logging mechanism, one log file is created corresponding to each file being transferred. For example, consider $File_A$ 
that needs to be transferred to another data center for analysis, LADS data transfer tool, segments the file into N objects (or) blocks. Upon successful completion of block K ($B_k$), the file logger mechanism will write to the log file that $B_k$ was successfully transferred and written to the parallel file system (PFS) at sink end. Due to any fault, if it needs to restart the transfer, LADS first searches for completed blocks from the corresponding log file and builds list of the blocks that were not transferred to the sink end. Then it starts sending only these blocks. Once all the blocks, corresponding to one logical file being transferred, have been successfully transferred and written to the PFS, the log file will be deleted. 

This mechanism is easy to implement. As the log is maintained for a single file, the search overhead for retrieving the completed block information should consume much less time because the log file size is relatively smaller than other 
mechanisms such as transaction or universal log mechanisms (to be explained later). However, as each and every file is associated with one log file, an increase in the number of files in the dataset will have direct impact on the number of logger files created. 
To avoid this, light-weight logging mechanism is implemented. Using light-weight logging, log files are created only when the first object of the new file is transferred successfully and deleted upon completion of the transfer.

\subsubsection{Transaction Logger}
In contrast to the file logger mechanism, transaction logger makes use of one log file for one transaction. Size of the transaction can be configurable depending on the total dataset size. 



Using single log file for maintaining the completed objects information of multiple logical files, would need methods to differentiate the data of one file from another. To achieve this, an index file is used. Index file contains the information of the file being transferred. Each line in index file looks like,
\begin{center}
	{\textit{[LogFileName, FileName, TotalBlocks, Offset, Data\_Length]}} \\
\end{center}
where,
\begin{center}	
	\begin{tabular}{l}	
		\textit{LogFileName}	$\,\to\,$	Transaction logger file name\\
		\textit{FileName}		$\,\to\,$	Name of the file being transferred\\
		\textit{TotalBlocks}	$\,\to\,$	Number of Blocks\\
		\textit{Offset}			$\,\to\,$	Offset in the logger file\\
		\textit{Data\_Length}	$\,\to\,$	Length of data in log file\\		
	\end{tabular}
\end{center}		

With this logger mechanism, the search overhead for retrieving the completed block information is much similar to file logger mechanism. However, computational complexity would be higher than file logger mechanism due to logging multiple files data rather than single file data.

\subsubsection{Universal Logger}
Universal logger mechanism is much similar to transaction logger in the way the completed blocks information is logged in the logger file. However, in contrast to transaction logger, universal logger makes use of single logger file corresponding to all the files to be transferred from one source node. As single log file is used for maintaining the completed objects information of multiple logical files, would need methods to differentiate the data of one file from another. To achieve this, an index file is used. Index file contains the information of the file being transferred. Each line in index file looks like,
\begin{center}
	\textit{[FileName, TotalBlocks, Offset, Data\_Length]}\\
\end{center}
where,
\begin{center}	
	\begin{tabular}{l}	
		\textit{FileName}		$\,\to\,$	Name of the file being transferred\\
		\textit{TotalBlocks}	$\,\to\,$	Number of Blocks\\
		\textit{Offset}			$\,\to\,$	Offset in the logger file\\
		\textit{Data\_Length}	$\,\to\,$	Length of data in log file\\		
	\end{tabular}
\end{center}		

With universal logger mechanism, search overhead and computational complexity for logging the completed blocks information are much similar to transaction logger.


\subsection{Object based logging Methods}
\label{sec:methods}
The logger mechanisms described above are analyzed with different object based logging methods. These methods vary on how log information is stored.

\begin{itemize}
	
	\item
	{Char type:} 
	The block number to be populated in the log file will be converted to string first and then written to the file. 
	\item
	{Encoding type:} 
	Successful block {information with the char type} 
	will be encoded using a Variable Length Datatype (VLD) library written by one of the authors.
	\item
	{Int type:} 
	Successful block will be written to the file using integer data. 
	\item
	{Binary type:} 
	Before writing to the file, block number is first converted to binary format.
	Assuming any file under consideration is not segmented more than
	$2^{32}$
	number of blocks, currently we are using 32-bit binary representation. 
	\item
	{Bit binary:} 
	{Each} bit is used to represent one block. For example, transferring block $K$ has been completed successfully and considering $N$-bit approach, we can represent that block in this method 
	by calculating the array index ($i$) and bit position ($j$) as,
	$Array_i$ = K / N  and $Bit_j$ = K mod N.
	Setting the bit in the $Bit_{Position}$ of the corresponding index, $Array_{index}$ will indicate the completion of the transfer of that particular block. In this method, 
	{we compare} the space and execution time by using both 8-bit and 64-bit.
	Pseudo code for bit-binary method of logging is as shown in Algorithm~\ref{algo:methods}.

	\begin{algorithm}[t!]
		\caption{Bit Binary Method of Logging (N=8 or N=64)}
		\begin{algorithmic}[1]	
			\Procedure{BITBINARY}{A, N}
			\State buff $<$$-$ ReadFromFile;
			\State ArrayIndex = A/N;
			\State Bit$_{Pos}$ = A\%N;	
			\State buff[ArrayIndex] = buff[ArrayIndex] $|$ ($1$ $<<$ Bit$_{Pos}$)
			\State WritetoFile $<$$-$ buff;
			\EndProcedure
		\end{algorithmic}
		\label{algo:methods}
	\end{algorithm}

\end{itemize}

\section{Fault-Tolerance Design With LADS}
\label{sec:Fault}

Fault tolerance support for Layout-Aware Data scheduler (FT-LADS) is motivated to answer a simple question: how can we improve the LADS data transfer performance in case of software, hardware or common communication errors. 

In this section, we describe the design and architectural changes incorporated in LADS to support fault tolerance. 

\subsection{Sequence Flow of FT-LADS}

\begin{figure}[!t]
	\begin{center}
		\includegraphics[width=0.62\textwidth]{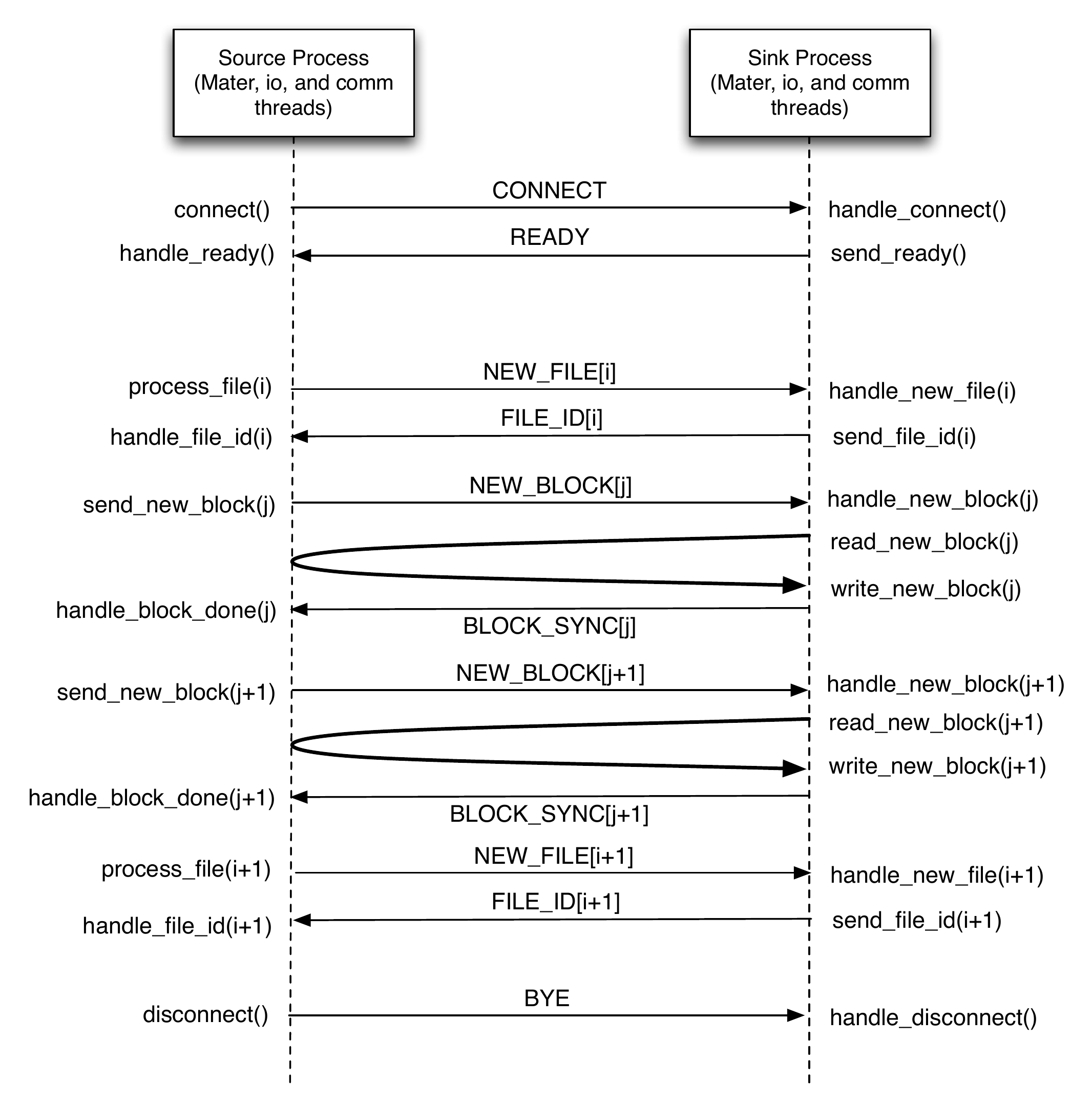}
		\caption{Communication Protocol between source and sink }
		\label{fig:ftlads_proto}
	\end{center}	
\end{figure}

Proposed FT-LADS communication protocol between the source and the sink end points is as shown in Figure~\ref{fig:ftlads_proto}. The BLOCK\_DONE message in LADS has been modified to BLOCK\_SYNC message to handle PFS write failures as listed in Listing~\ref{list1}. Upon receiving the BLOCK\_SYNC message, based on the synchronous or asynchronous logging method, the source \textit{comm} thread either writes the completed block information to the FT logger file directly or enqueues the request on the wait queue in the logger thread. 
In case of synchronous logging, the completed objects information is populated to the FT logger file in the context of the \textit{comm} thread. Whereas, in case of asynchronous logging, a different \textit{logger} thread is used for logging the completed objects information to the logger file. In both cases, we implemented and evaluated the performance and found no difference between the two methods. Therefore, we present only synchronous logging mechanism.

\begin{figure}[!t]
	\begin{center}
		\includegraphics[width=0.95\textwidth]{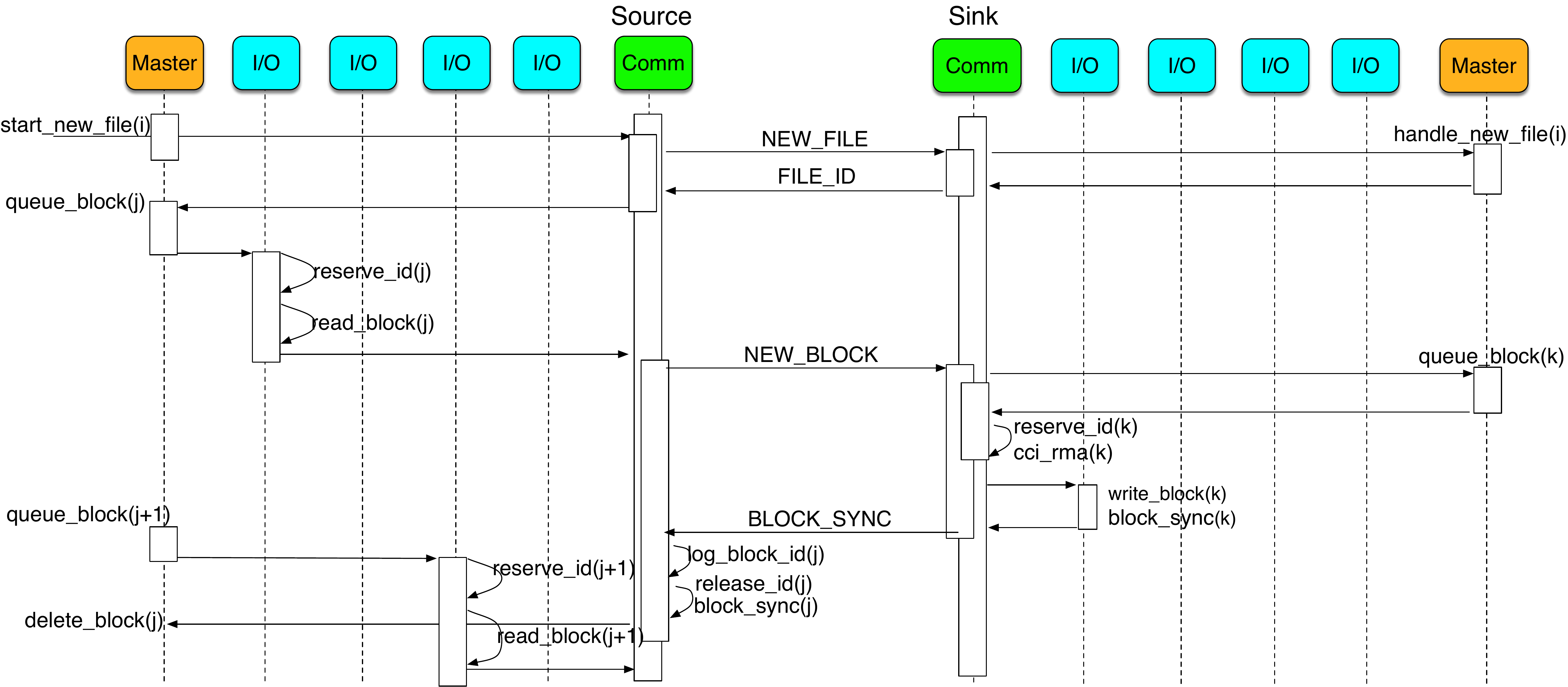}
		\caption{Fault tolerant LADS data transfer sequence diagram}
		\label{fig:ftlads_seq}
	\end{center}	
\end{figure}

The flow of data transfer in fault tolerant LADS between the source and the sink endpoints is shown in Figure~\ref{fig:ftlads_seq}. For each file in the target dataset, the source \textit{master} thread generates a NEW\_FILE request and enqueues the request on the work queue in the \textit{comm} thread. The \textit{comm} thread dequeues the request and transfers it to the sink end using CCI interface. At the sink end, the \textit{comm} thread receives the NEW\_FILE request and enqueues the request to the \textit{master} thread's work queue and wakes it up. Based on the target file information in the request, the \textit{master} thread opens the file and adds the file descriptor to the FILE\_ID request and then enqueues the request to the \textit{comm} thread's work queue. The \textit{comm} thread dequeues the request and sends it to the source. On receiving the FILE\_ID request, the \textit{comm} thread 
enqueues the request on the \textit{master} thread's wait queue and wakes it up. The \textit{master} thread splits the file as per object size and generates the NEW\_BLOCK requests and enqueues the requests on the \textit{I/O} thread wait queue and wakes it up.  An \textit{I/O} thread first reserves a buffer registered with the CCI for RMA. It then determines which OST queue it should access and then dequeues the first NEW\_BLOCK request. It uses pread() to read the data into the RMA buffer. When the read completes, it enqueues the request on the \textit{comm} thread's work queue. The \textit{comm} thread dequeues the request and transfers it to the sink. At the sink, the \textit{comm} thread receives the request and attempts to reserve the RMA buffer. If successful, it initiates an RMA read operation of the data. If not, it enqueues the request on the \textit{master} thread's work queue and wakes the \textit{master} thread. The \textit{master} thread will sleep on the RMA buffer's wait queue until a buffer is released. Once the buffer is available, the request is placed on the \textit{comm} thread's queue, which will issue RMA read operation. Upon completing the RMA read operation, the sink's \textit{comm} thread determines the appropriate OST by the object's file offset and queues it on the OST$'$s work queue. It then wakes up an \textit{I/O} thread. The \textit{I/O} thread looks for the next OST to service and dequeues a request and then calls pwrite() to write the data to the disk. When the write is completed, it releases the RMA buffer so that the \textit{comm} thread can initiate another RMA read operation and also sends the BLOCK\_SYNC request to the source. Upon receiving the BLOCK\_SYNC message, the source \textit{comm} thread, based on synchronous or asynchronous logging method, logs the completed block information to the FT file or enqueues the request to the \textit{logger} thread wait queue respectively. 
Logging method will vary based on the logger mechanism and the method options. This process is carried on till all objects of the data are successfully transferred to the sink (or) till there is any fault.

\vspace{-0.2in}
\begin{center}
	\begin{lstlisting}[float, style=CStyle, label=list1]
	typedef enum msg_type {
	CONNECT = 0,	//Connect Request
	NEW_FILE,		//New File request
	FILE_ID,		//Sink File ID.
	NEW_BLOCK,		//Ready for RMA Read
	BLOCK_SYNC,		//Sync with Sink PFS
	BYE,			//ready to disconnect
	FILE_CLOSE,		//file close
	} msg_type_t;	
	\end{lstlisting}
\end{center}

\subsection{Resuming Failed Transfers}
A fault tolerant design enables LADS to resume with the current data transfer from the same point as it was interrupted, upon recovery. When transfer is initiated,  based on the selected object logger mechanism and method as described in Section~\ref{sec:logger}, logger file will be created in {\textit{ftlads}} subdirectory under user home directory. If data transfer is initiated by enabling fault tolerance option, this subdirectory will be created automatically. The actual log file name under this subdirectory will be varied based on the selected object logger mechanism.

This section describes the steps performed before and after fault to resume the transfers upon recovery from fault.


\subsubsection{Before Fault}
Upon scheduling the data transfer, the source creates a NEW\_FILE request with the current file's metadata and sends the request to the sink. Based on the NEW\_FILE request information, the sink opens a file, creates a FILE\_ID request with sink end file descriptor and sends it to the source. On receiving the FILE\_ID request, the source schedules all the objects of the file for transfer. Upon successful transfer and writing to the PFS, a BLOCK\_SYNC message will be sent from the sink to the source. On receiving the  BLOCK\_SYNC message, the source writes the completed object information to the FT logger file. If all the objects are successfully transferred, then the FT log entry corresponding to that file is deleted.

\subsubsection{After Fault}
On resuming the transfer, the source creates a NEW\_FILE request with the current file's metadata and sends the request to the sink. On receiving the NEW\_FILE request, the sink checks if the file already exists and the file's metadata is matching with the source file's metadata. If matching, the file from the list of files to be transferred is skipped. If the file does not exist (or) the metadata is not matching, the sink creates a FILE\_ID request and sends it to the source. Upon receiving the FILE\_ID request, the source checks if the FT logger file corresponding to the file exists in the FT logger directory. If exist, the objects that were successfully transferred are retrieved. Then, the source builds the object list by excluding already completed objects and then schedules the transfer.

\section{Evaluation}
\label{sec:expr}

For the evaluation of FT-LADS, we have created a simulation environment 
where 
we have induced hardware faults during data transfer. First, we evaluate the fault tolerance overhead on LADS by showing the results of FT-LADS without fault. Then we explore the effectiveness of FT-LADS by comparing the recovery time overhead in FT-LADS with bbcp by varying fault points. All our experiments were conducted under similar conditions.

\subsection{Experimental Environment}

{\bf Implementation:}
FT-LADS, which is based on server-client model, has been implemented using $6K$ lines (including both LADS and fault tolerance implementation) of C code using Pthreads. We have used CCI, an open-source communication interface, which can be downloaded from CCI-Forum \cite{cci-forum}. 


{\bf Test-bed:}
For our experiments, we used a private testbed with two nodes (source and sink) connected by InfiniBand (IB). The nodes use the IB network to communicate with each other. 
We have used Intel(R) Xeon(R) CPU E5-2650 v4 @ 2.20GHz servers with $24$ cores and 128 GB DRAM. Both source and sink hosts are running with Linux kernel 3.10.0-514.21.1. 
Also, the source and the sink nodes have separate Lustre file systems 2.9.0 \cite{lustre-internals} with one OSS and 11 OSTs, mounted over 1 TB drives each. 
By default, our Lustre file system configuration includes stripe count of one with stripe size of $1$ MB. To fairly evaluate our implementation, we have ensured that the storage server bandwidth is not over-provisioned with respect to the network bandwidth between those source and sink servers (i.e., the network would not be the bottleneck).

{\bf Workloads:}
It is observed that 
90.35 percent
of the files are less than 4 MB and 86.76 percent are less than
1 MB~\cite{kim:fast15}~\cite{lads17}. Less than 10 percent of the files are greater than 4 MB
whereas the larger files occupy most of the file system
space. For the purpose of evaluation, we had used two groups of files with different sizes; one for small workloads with 10,000 1 MB files, and the other for big workloads with 100 1 GB files. For evaluation, we have pre-populated source host's file system with big and small workloads where file's stripe count is 1 and size is 1MB. 

{\bf Configuration:}
Experimental results presented in LADS \cite{kim:fast15}~\cite{lads17}, suggest that LADS data transfer performance increases linearly with the number of I/O threads. To have an optimal evaluation environment, in all our experiments, we have configured FT-LADS to use $4$ I/O threads, $1$ master thread, and $1$ comm thread. 

In case of transaction logger, we have considered $4$ files in one transaction. If the transaction size is set to 1, then the transaction logger is same as the File logger mechanism, as each and every file will be associated with one log file. If the transaction size is set to maximum, then the transaction logger is same as the Universal logger. So for our evaluations, we have used intermediate size as transaction size.

All the experiments were done by utilizing a large, fixed amount of DRAM used as RMA buffers at both the source and the sink. Our current implementation makes use of max. 256 MB of DRAM at both source and sink. We have run multiple iterations of all the experiments and shown average as bar graph. Also, 99\% confidence intervals are shown in error bar, wherever is needed.


{\bf Recovery Time:}
As there is no direct method of evaluating the recovery time, we have estimated the recovery time of failed transfers as below.

\begin{equation}\label{equ1}
ER_t =  TBF_t + TAF_t - TT_t
\end{equation}
\begin{center}
	\begin{tabular}{l}
		where, \\
		ER$_{t}$	Estimated Recovery Time \\
		TBF$_{t}$	Time consumed before fault\\
		TAF$_{t}$	Time consumed after fault \\
		TT$_{t}$	Time consumed with no fault \\
	\end{tabular}
\end{center}


\subsection{Performance comparison with LADS}
\label{subsec:perf_comp}

One of the major objectives while designing the object based fault tolerance mechanisms, is to minimize the object based FT overhead on LADS data transfer time. 
In this section, we present the evaluation results of different object based FT mechanisms and methods described earlier (Section ~\ref{sec:logger}). For evaluating the data transfer rate and computational overhead of FT-LADS, we have used total time to transfer, CPU load and memory usage as performance factors.

\begin{figure*}[!t]
	\centering
	\begin{tabular}{@{}ccccc@{}ccccc@{}ccccc@{}cccc@{}c@{}}
		\includegraphics[width=0.3\textwidth]{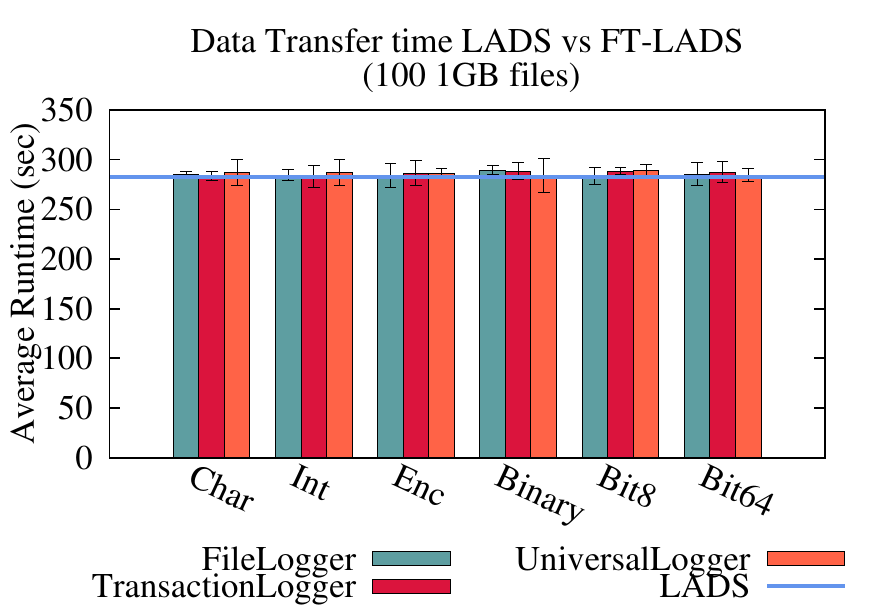}&
		\includegraphics[width=0.3\textwidth]{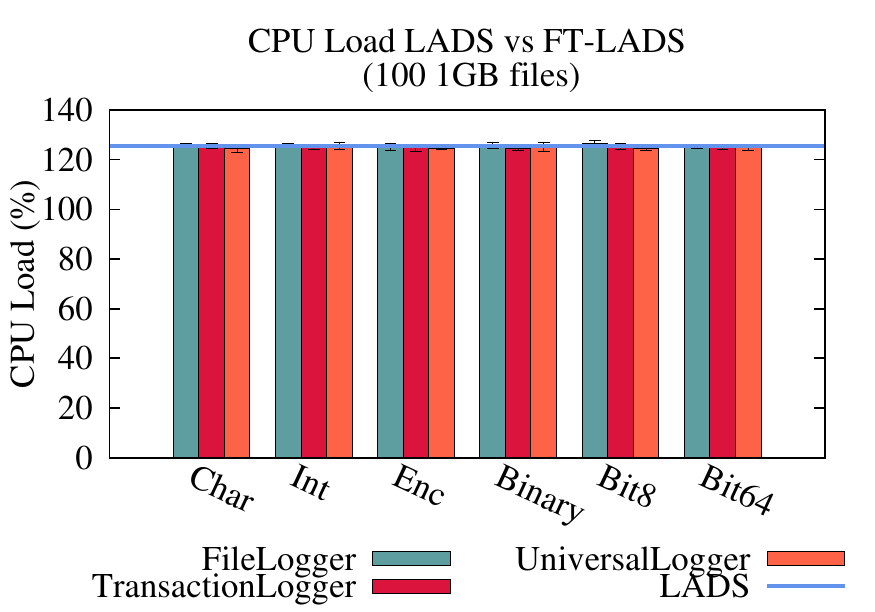}&
		\includegraphics[width=0.3\textwidth]{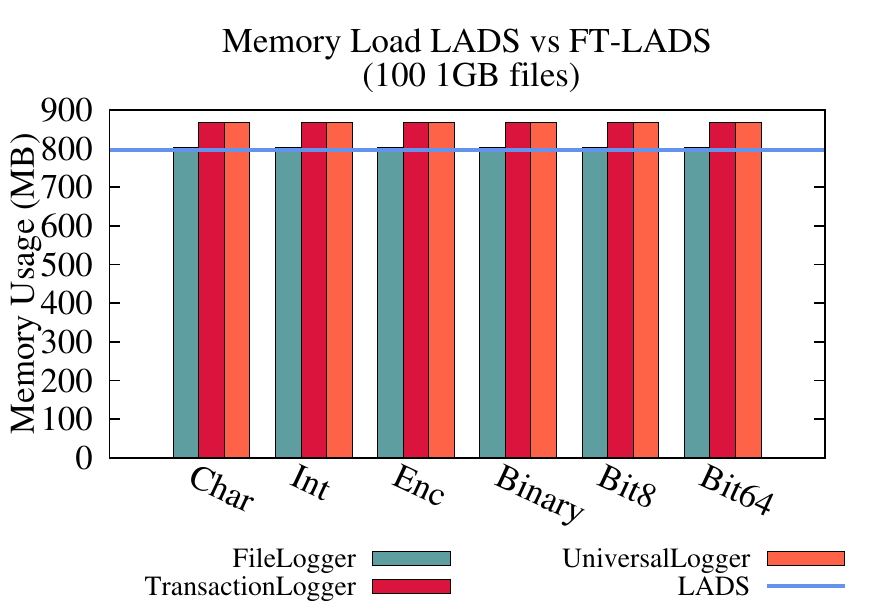} \\
		\small (a) Data transfer Time &
		\small (b) CPU Load &
		\small (c) Memory usage \\
	\end{tabular}
	\caption{\small Performance comparison of LADS and FT-LADS for Big Workloads. The 99\% confidence intervals are shown in error bar.}
	\label{fig:perf_big}
	\vspace{0.1in}
\end{figure*}

\begin{figure*}[!t]
	\centering
	\begin{tabular}{@{}ccccc@{}ccccc@{}ccccc@{}cccc@{}c@{}}
		\includegraphics[width=0.3\textwidth]{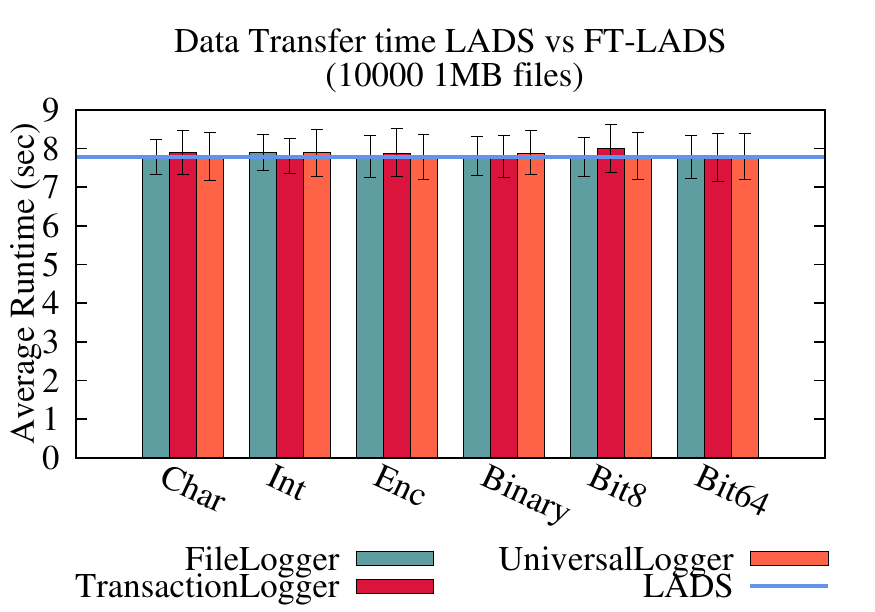} &
		\includegraphics[width=0.3\textwidth]{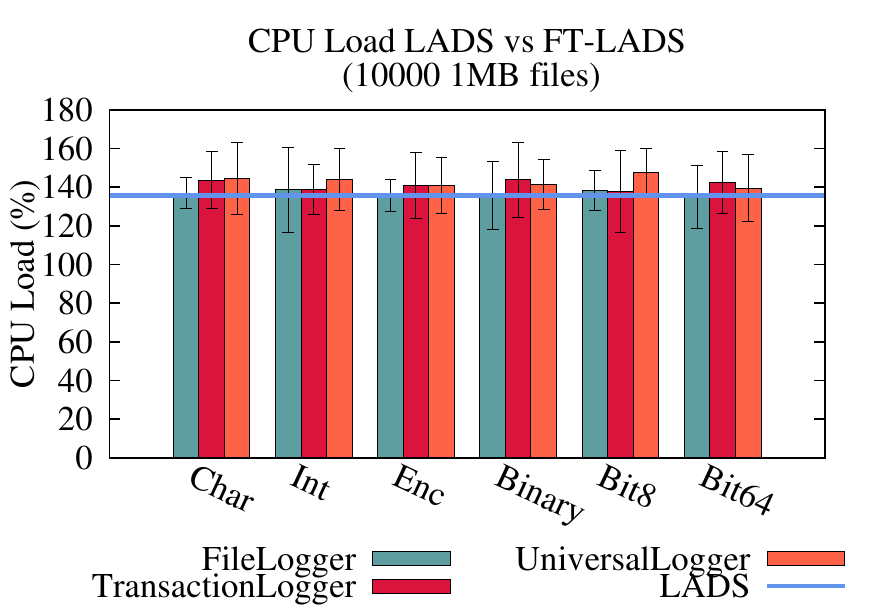} &
		\includegraphics[width=0.3\textwidth]{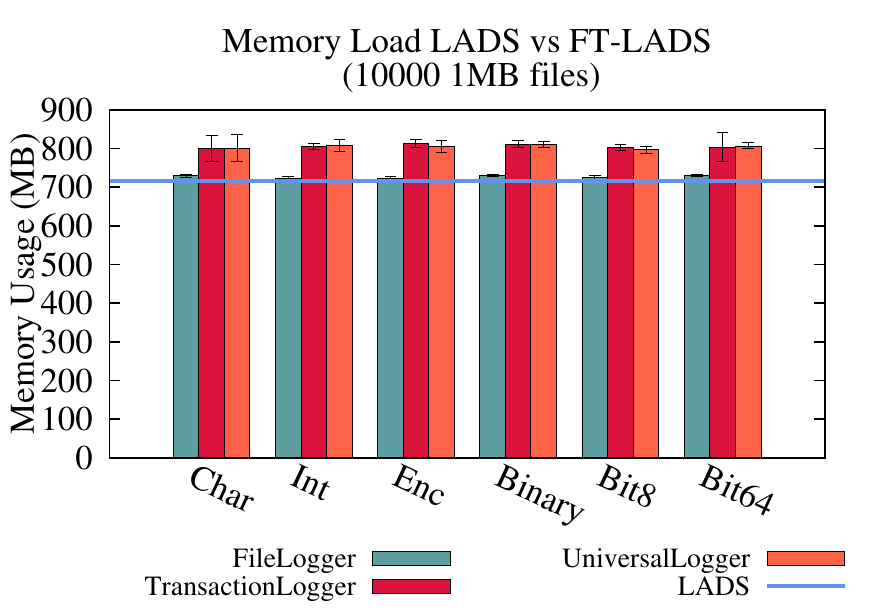}\\
		\small (a) Data transfer Time &
		\small (b) CPU Load &
		\small (c) Memory usage \\
	\end{tabular}
	\caption{\small Performance comparison of LADS and FT-LADS for Small Workloads. The 99\% confidence intervals are shown in error bar.}
	\label{fig:perf_small}
\end{figure*}

Figure~\ref{fig:perf_big} and~\ref{fig:perf_small} show the performance comparison between LADS and FT-LADS. In these figures, the proposed object based mechanisms are represented using bar graph, whereas 
line is used to represent LADS. 

Figure~\ref{fig:perf_big}(a) and Figure~\ref{fig:perf_small}(a) depict the total time consumed for transferring the big workloads and the small workloads respectively. From Figure~\ref{fig:perf_big}(a) and Figure~\ref{fig:perf_small}(a), we can clearly observe that all the proposed FT mechanisms have negligible impact on the overall data transfer time. With this, we can conclude that the proposed FT mechanisms have no impact on the overall data transfer rate, as data transfer time is inversely proportional to the data transfer rate.

Total CPU load during data transfer is another important design aspect, while designing FT support with LADS. 
Figure~\ref{fig:perf_big}(b) and Figure~\ref{fig:perf_small}(b) depict the CPU load while processing the data transfer. From 
these figures, we can observe that, there is no significant impact on the total CPU load with FT support, compared with LADS.

Figure~\ref{fig:perf_big}(c) and Figure~\ref{fig:perf_small}(c) represent memory load comparison of proposed FT mechanisms with LADS. From these figures, we can clearly observe that, with file logger mechanism, there is no impact on the memory load, whereas with other mechanisms, we can see an increase in the memory load. In case of File logger mechanism, we simply write the completed object information to the corresponding FT logger file and there is no additional data structures which will be used to save the intermediate data. Whereas, in case of transaction and universal logger mechanisms, completed objects information of multiple files need to be logged interleavingly as single logger file is used. This will increase the recovery time upon fault. To optimize the recovery time, completed objects information of all files are maintained internally as a list before actually logging into the logger file. Due to the use of intermediate data structure, the total memory used by transaction and universal mechanisms is higher than those of File logger and LADS.

From the 99\% confidence intervals which are shown as error bar in Figure~\ref{fig:perf_big} and~\ref{fig:perf_small}, we can observe that there is a lot of variability for small workloads with respect to data transfer time, CPU load and memory load.  This variability might be due to the file management overhead of the file system, as the number of files to be transferred is much higher in small workloads.

As shown in Figure~\ref{fig:perf_big} and~\ref{fig:perf_small}, the performance is not affected by the FT methods (Char, Int, Enc, Binary, Bit8 and Bit64) used for both big and small workloads. With this, we can conclude that all the proposed object based FT mechanisms and methods have minimum to negligible performance overhead compared to LADS and the file logger is the most lightweight mechanism with minimal overhead among the proposed FT mechanisms.



\subsection{Object based Logger Methods Space Analysis}
Another important aspect while designing the FT-LADS is the amount of space occupied by the logger files during data transfer. To optimize the log space occupied, as mentioned in Section~\ref{sec:methods}, we have proposed different logging methods. In this section, we compare the space occupied by different logger methods.

\begin{figure}[!t]
	\centering
	\includegraphics[width=0.4\textwidth]{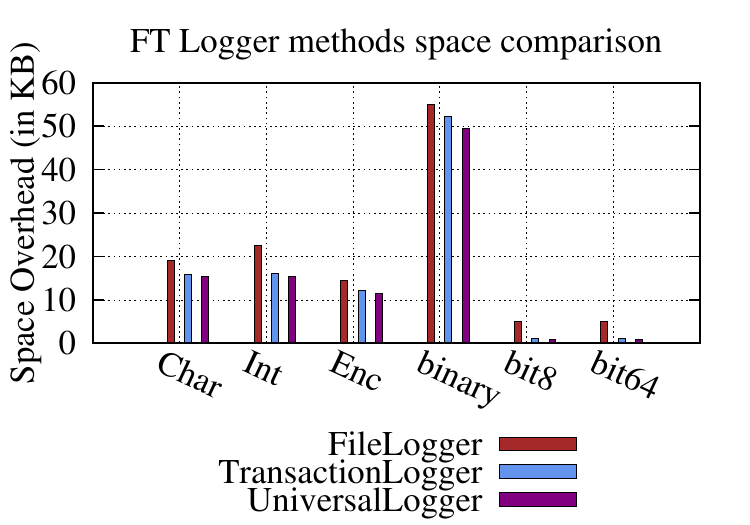} \\
	\caption{\small FT Logger Methods space overhead}
	\label{fig:ft_space}
\end{figure}

Figure~\ref{fig:ft_space} depicts the space overhead of all the proposed logging methods for all the object based fault tolerance mechanisms. From the figure, it is evident that bitbinary (Bit8 and Bit64) method is the most effective among all the logger methods due to its low space overhead. This is expected as each object is represented with one bit. Though other logging methods have relatively higher space overhead than bitbinary method, the overhead is quite negligible which is in the order of few KB. 

As mentioned in section~\ref{subsec:perf_comp}, all the proposed FT methods have negligible performance overhead among each other. With this, we can conclude that Bit8 and Bit64 FT methods are recommended with respect to space overhead with the proposed object based FT mechanisms. 

From Figure~\ref{fig:ft_space}, we can also observe that among all the proposed FT methods, Universal logger mechanism has minimal space overhead when compared with other mechanisms. But considering file logger mechanism's minimal performance overhead, we can conclude that file logger FT mechanism with bitbinary (Bit8 and Bit64) FT methods is the most suitable object based fault tolerance mechanism.

\subsection{Recovery time analysis}
\label{rec_ana}

Minimizing the recovery time upon resuming from fault is one of the major objectives in our FT-LADS design. In this section, we have evaluated the FT-LADS recovery time for small and big workloads and compared it against bbcp data transfer tool. For effective evaluation of recovery time of proposed fault tolerance methods, we created a simulation environment in which we generate faults after transferring 20\%, 40\%, 60\%, 80\% of total data size. 
As faults can occur at any end of the transfer, we can simulate the faults at either source or sink. However, for the purpose of our experiments, we have executed this simulation in the source end. 
Using the experimental environment, described in Section~\ref{sec:expr}, we measured the recovery time for both bbcp and FT-LADS. 
On these hosts, LADS uses CCI's Verbs transport, which natively uses the underlying InfiniBand interconnect. Whereas, bbcp uses the IPoIB interface which supports traditional sockets. 

In LADS, varying the number of I/O threads maximizes CPU utilization on the data transfer node. 
However, bbcp uses configurable window size and multiple streams to improve the performance. Based on the experimental results presented in LADS \cite{kim:fast15}~\cite{lads17}, LADS data transfer performance increases linearly with the number of I/O threads. Whereas, bbcp has less impact while increasing the number of tcp streams. For fair performance comparison between the two, we have configured FT-LADS to use $4$ I/O threads and bbcp to use $2$ tcp streams with window size of 8MB.
Our experiments are designed to calculate the transfer time before and after fault. Based on these times and using Equation~\ref{equ1}, we estimated the recovery time.

\begin{figure*}[!t]
	\centering
	\begin{tabular}{@{}cccc@{}c@{}c@{}c@{}c@{}}
		\includegraphics[width=0.35\textwidth]{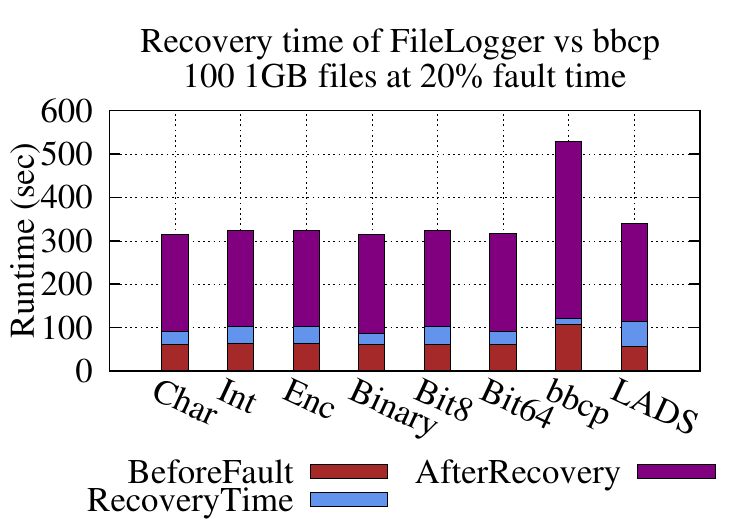}&
		\includegraphics[width=0.35\textwidth]{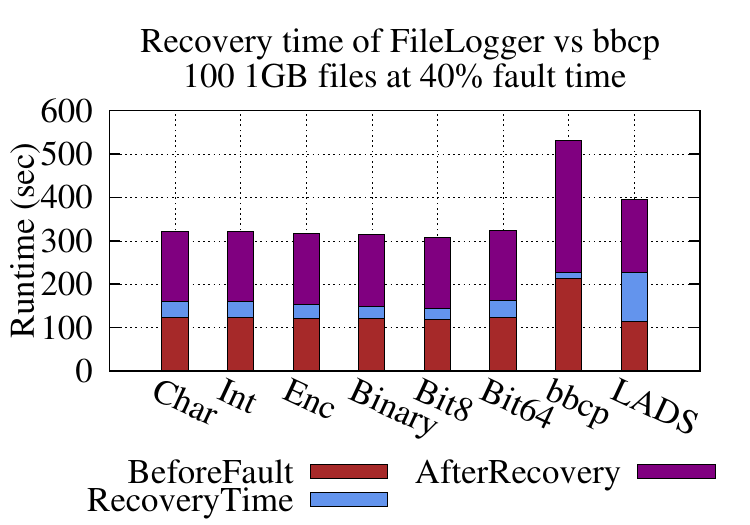}&\\
		\small (a) Big loads 20\% fault time  &
		\small (b) Big loads 40\% fault time  &\\				
		\includegraphics[width=0.35\textwidth]{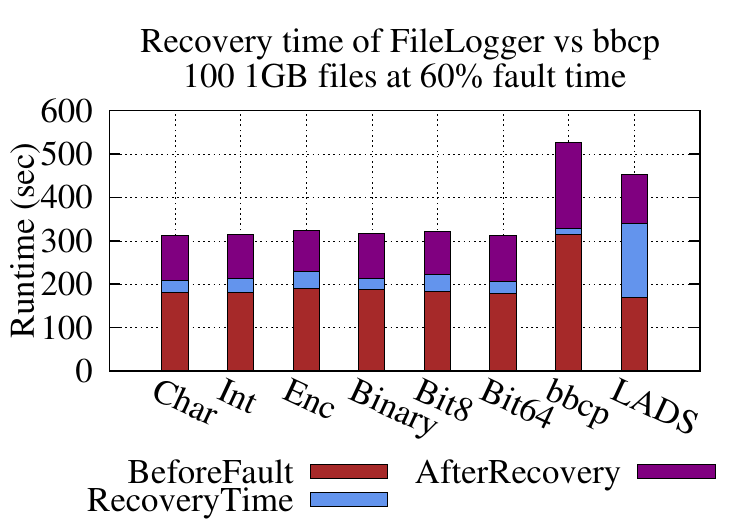}&
		\includegraphics[width=0.35\textwidth]{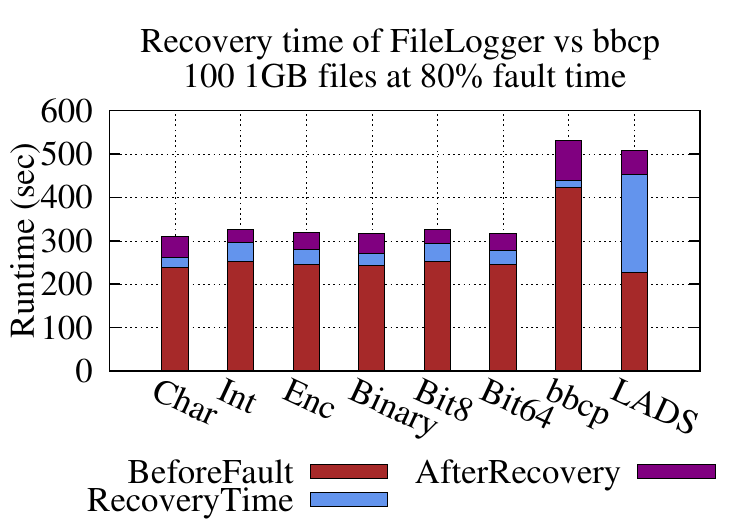}\\
		\small (c) Big loads 60\% fault time  &
		\small (d) Big loads 80\% fault time  \\
	\end{tabular}
	
	\caption{\small Recovery time analysis of FileLogger at varying fault timing for big workloads.
	}
	\label{fig:bbcpvsFL_big}
\end{figure*}

\begin{figure*}[!t]
	\centering
	\begin{tabular}{@{}cccc@{}c@{}c@{}c@{}c@{}}
		\includegraphics[width=0.35\textwidth]{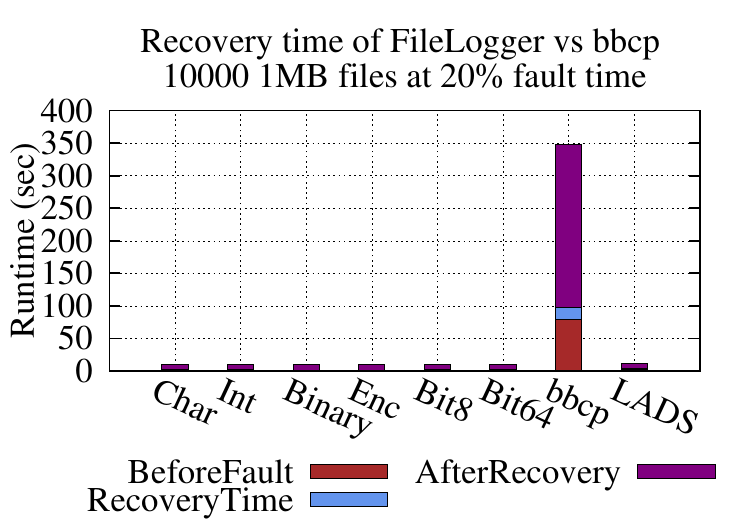}&
		\includegraphics[width=0.35\textwidth]{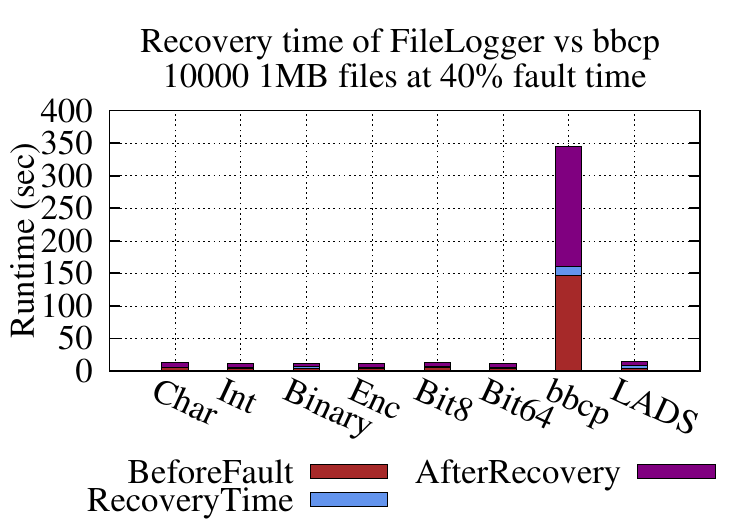}&\\
		\small (a) Small loads 20\% fault time  &
		\small (b) Small loads 40\% fault time  &\\		
		\includegraphics[width=0.35\textwidth]{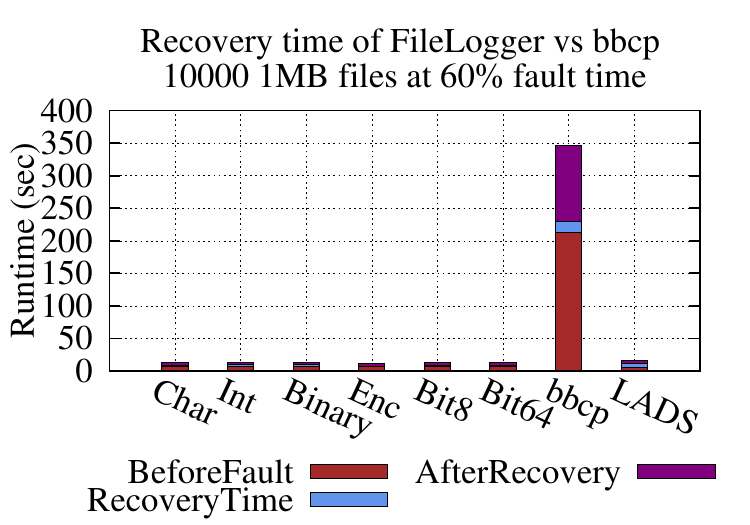}&
		\includegraphics[width=0.35\textwidth]{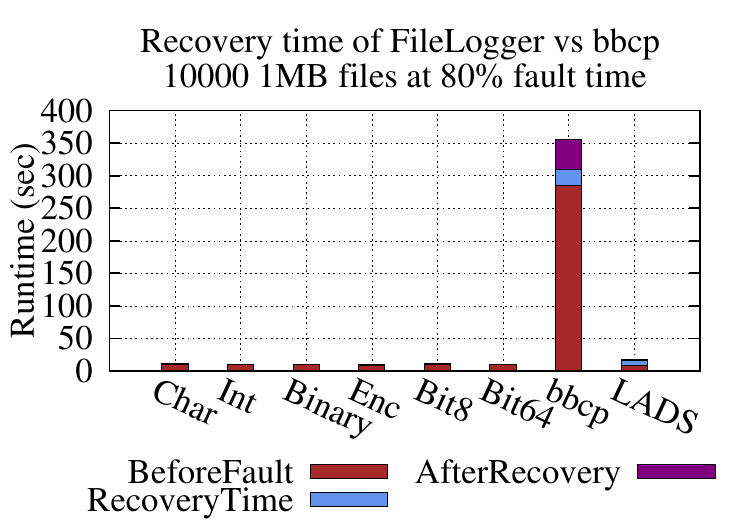}\\
		\small (c) Small loads 60\% fault time  &
		\small (d) Small loads 80\% fault time  \\
	\end{tabular}
	
	\caption{\small Recovery time analysis of FileLogger at varying fault timing for small workloads.
	}
	\label{fig:bbcpvsFL_small}
\end{figure*}

Recovery times with all object based fault tolerance mechanism and methods are compared with that of bbcp data transfer tool. Where we set LADS recovery time as the baseline for our experiments. As resume operation is not supported in LADS, LADS has to transfer all the objects of the dataset upon resuming from faults. From the experimental results shown in Figure~\ref{fig:bbcpvsFL_big},~\ref{fig:bbcpvsFL_small} and ~\ref{fig:ft_rec}, the later the fault occurs, the higher the recovery time is. 
Our aim is to minimize the impact of recovery time on the fault point. As per our logging mechanism, we delete the log file entries of the logical files, which are successfully transferred to the sink end. Due to this at any point of time, we are left with only those files which are currently being progressed. 
The amount of logs to be parsed to retrieve the objects which are successfully synchronized at the sink end PFS will not depend on the fault point. 

Recovery time of File logger mechanism at varying fault points for both big and small workloads is as shown in Figure~\ref{fig:bbcpvsFL_big} and~\ref{fig:bbcpvsFL_small}. For other Transactional and Universal FT mechanisms, similar results were observed. We only show the results for File logger in this paper.


\subsubsection{Big Workloads}
\label{sec:bigwork}
In case of file logger mechanism, the recovery times for all fault tolerance methods exhibit similar recovery times irrespective of the fault points (Refer to Figure~\ref{fig:bbcpvsFL_big}). 
Though the recovery time is much lower than LADS, all the methods of file logger mechanism consume higher recovery times than bbcp. 
As bbcp FT is based on file offset, its recovery time is much less than that of file logger.
This is expected as in file logger mechanism, each logical file to be transferred is associated with one log file and while writing the logs to file, we just append the completed object index at the end of logger file. Due to this, while retrieving the completed object information, an additional search overhead is involved.

For transaction and universal loggers, the recovery time overhead of big workloads is negligible. This is also expected, as the completed objects information is sorted as per object index before writing to the logger file. 
As mentioned in Section~\ref{subsec:perf_comp}, we are using intermediate lists, which maintain the completed objects information of all files being transferred, by sorting based on object index.

\subsubsection{Small Workloads}
In contrast to the big loads, bbcp tool consumes much higher transfer time for smaller workloads than LADS. 
Due to this, the recovery time overhead of FT-LADS is not directly comparable with bbcp. For quantitative comparison, percentage of recovery time relative to each method is calculated. At all given fault points, bbcp exhibit 5\% to 7\% recovery time overhead. Whereas, all the proposed FT methods experience around 12\%-14\% overhead.


Our small workload consists of files whose size is of 1MB and this is matching our transfer unit size. Due to this, a file transfer state can be either completed or transferred upon recovery from fault. So, there won't be any log files which need to be parsed upon fault, and hence, proposed object based logger mechanisms just determine which files are already completed and start transferring the remaining files. As a result, we can conclude that with the proposed object based fault tolerance mechanisms, the recovery time overhead will not come into the picture.

\begin{figure}[!t]
	\centering
	\begin{tabular}{@{}cccc@{}c@{}}
		\includegraphics[width=0.4\textwidth]{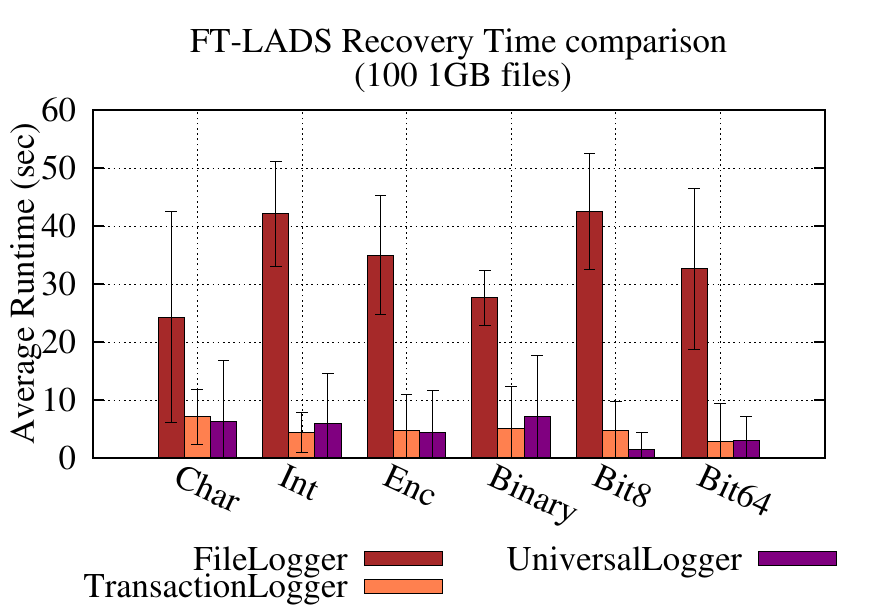} &
		\hspace{0.3in}		
		\includegraphics[width=0.4\textwidth]{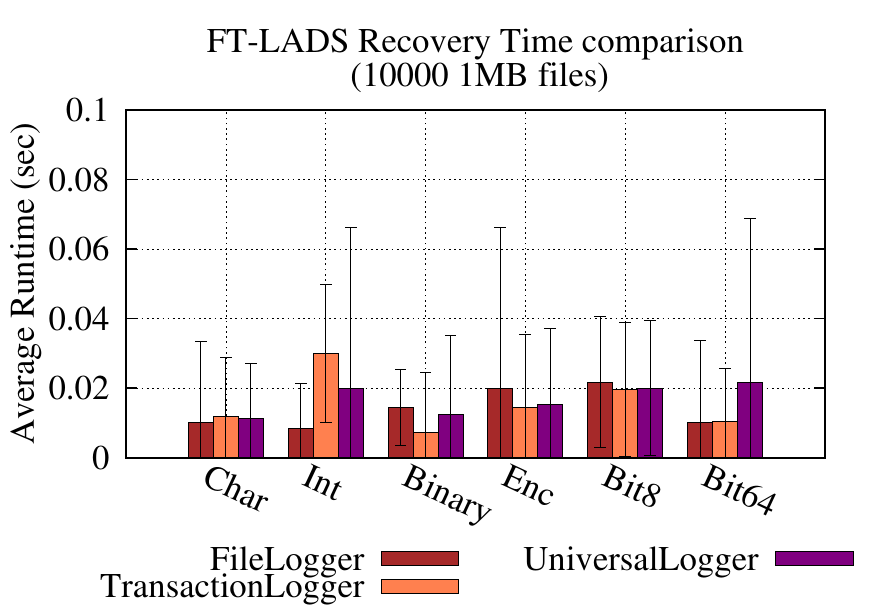} \\		
		\small (a) Big loads 80\% fault time  &
		\small (b) Small loads 80\% fault time  \\		
	\end{tabular}
	\caption{\small Recovery time analysis of FT Loggers at 80\% fault timing. The 99\% confidence intervals are shown in error bar.}
	\label{fig:ft_rec}
\end{figure}

In Figure~\ref{fig:ft_rec}, we have shown the recovery time comparison among the proposed fault tolerance mechanisms, considering 80\% fault point as a reference, for both big and small workloads. As shown in Figure~\ref{fig:ft_rec}, we can observe that for big workloads, file logger mechanism exhibits higher recovery time than other proposed FT mechanisms. Whereas, for small workloads, the recovery overheads for all mechanisms and methods are similar as shown in Figure ~\ref{fig:ft_rec} (b).

From Figure ~\ref{fig:ft_rec} (a) and ~\ref{fig:ft_rec} (b), we can observe that Universal logger mechanism exhibits lower recovery times upon fault. Also, among all the FT methods, bitbinary methods (Bit8 and Bit64) have minimal recovery overhead compared with the other FT methods.


Based on our evaluation results, file logger mechanism shows minimal impact on the performance while logging the completed objects information. Whereas, universal logger is superior to other mechanisms with respect to recovery times upon fault. Also, among the proposed FT methods, bitbinary methods (Bit8 and Bit64) have minimal space overhead and prove to have comparably lower recovery times among all the proposed FT mechanisms. Hence, conjugating LADS with universal object based FT mechanism and bitbinary FT methods will improve data transfer performance in faulty environments.

\section{Related Work} 
\label{sec:related}


To meet the needs of big data transfers, prior studies have performed on the design and implementation of bulk data movement frameworks ~\cite{Allcock:2005:GSG:1105760.1105819, bbcp, xdd:brad, sc12:rftp, Ren:sc13, Subramoni:2010:HPD:1844765.1845179, 13:vallee:ndm, DBLP:journals/tpds/LiRYJ17}. GridFTP~\cite{Allcock:2005:GSG:1105760.1105819}, which is an extended version of the standard File Transfer Protocol (FTP), provides high speed, reliable, and secure data transfer. The striping feature in GridFTP enables the support for multi-host to multi-host transfers. But this tool does not try to schedule the data transfer based on the underlying object locations. \textit{bbcp}~\cite{bbcp} is another data transfer tool which uses multiple streams for transferring large datasets. It uses a file based approach, which transfers the whole file data sequentially. 
XDD~\cite{xdd:brad} optimizes the disk I/O performance by enabling file access with direct I/Os and using multiple threads for parallelism, and varying file offset ordering to improve I/O access times. RAMSYS~\cite{DBLP:journals/tpds/LiRYJ17}, a resource-aware high-speed data transfer software, utilizes a multi-stage end-to-end data transfer pipeline, where each stage is fully resource-driven and implements a flexible number of components using predefined functions, such as storage I/O, network communication, and request handling. RAMSYS relies on the asynchronous paradigm to maximize the concurrency of components and thereby offers improved scalability and resource utilization in modern multi-core systems. All these tools are useful for moving large data faster and secure from source host to remote host over the network, but none of them tries to schedule based on the underlying object locations because they do not consider storage contention. 



Another important aspect of these data movement frameworks is to resume the data transfer upon faults during data transfer. GridFTP tool supports fault tolerance using restart markers (checkpoints). While transferring data, GridFTP server automatically sends restart markers to the client. If the transfer has a fault, the client may restart the transfer by providing the markers received. The server will restart the transfer from the point where it left off based on the markers. GridFTP's Reliable File Transfer (RFT) service provides an interface to write the restart markers to a database so that it can survive a local fault. \textit{bbcp} tool employs fault tolerance mechanism based on checkpoint record. Upon initiating a new transfer, \textit{bbcp} tool checks if checkpoint record of file being transferred exists or not. If record does not exist, it checks the target file attributes like name, size, etc. If they are identical with the source file attributes, then \textit{bbcp} assumes that the file transfer completed successfully and skips the transfer. If file attributes are different, then it initiates a new transfer by creating a checkpoint record and transmit all the source bytes to the target. Upon successful completion, it erases the checkpoint record. If checkpoint record exists, then it resumes the transfer by appending all untransmitted bytes to the target. XDD and RAMSYS tools did not implement fault tolerance, because they did not know which data needs to be transmitted upon fault.

As all the aforementioned bulk data movement frameworks transfer the logical file data sequentially, it is possible to resume transfers using checkpoint based restart marker or offset record. Checkpoint based fault tolerance methods are light-weight and also possible to resume transfer from restart marker or offset record without delay.

Our work focuses on entirely different scenario from the prior fault tolerance studies. Our work focuses on supporting resume functionality upon fault when the workload is transferred as objects rather than files, by exploiting the underlying storage architecture. Since a logical file is striped over multiple OSTs, it is possible to transfer one logical file's objects in random order. While, above mentioned checkpoint based restart marker or offset record is not sufficient to resume the transfer upon fault, our work proposes novel methods to handle fault tolerance in object based big data transfers. 

In our proposed object based fault tolerance mechanisms, objects which are successfully written to sink PFS are marked as successful and we update the information of the object in the logger file. Upon successful completion of all the objects of one logical file, the log information corresponding to the file will be erased. If there is any fault during the transfer, the proposed mechanisms search for the completed objects and schedule only those objects which are not transferred previously. As in object based fault tolerance mechanism, it needs to log all the objects of a file. This involves access to asynchronous filesystem API which causes processing overhead. It also results in space overhead as all the objects information is logged to the logger file. It involves an additional overhead to retrieve the completed objects information from the logger file for resuming the transfer upon recovery from fault. Our solution proposes methods to overcome processing, space and recovery time overheads.


\section{Conclusion} 
\label{sec:conc}
LADS data transfer tool with its layout-aware and OST congestion-aware algorithms, outperforms existing data transfer tools with respect to the data transfer rate. 
However, absence of fault tolerance support results in the data retransmission upon fault.
As LADS employs object level scheduling algorithms, objects of one logical file may be transferred out of order. 
Due to this, the fault tolerance mechanisms based on logging file offset, are not suitable for LADS. 
In this work, we have implemented object based fault tolerance mechanisms which can handle the out-of-order nature of object transmission. 
Depending on the number of logger files generated per dataset, we have proposed three different object logger mechanisms, \textit{File logger, Transaction logger and Universal logger}. 
Also in order to reduce the space overhead of logging, we have proposed six different fault tolerance methods: Char, Int, Enc, Binary, Bit8 and Bit64. We have evaluated the performance overhead of fault tolerance on LADS and concluded that proposed object based logging mechanisms do not impact the LADS data transfer performance. 
In order to evaluate the recovery time overhead of proposed object based fault tolerance mechanisms, we have created simulation environment to generate faults at 20\%, 40\%, 60\% and 80\% points of data transfer. 
From our evaluation results, we have observed that the recovery time in file logger mechanism exhibits $2$ times higher than bbcp, whereas the recovery times in transaction and universal logger mechanisms were considerably smaller than bbcp. 
To conclude, \textit{File logger} mechanism has minimal impact on logging the completed objects. 
Whereas, universal logger mechanism combined with bitbinary methods (Bit8 and Bit64) has a minimum overhead with respect to space and recovery times. Hence, with the proposed fault tolerance mechanisms, LADS can experience improved recovery performance upon fault during transfer.

\section*{References}

\bibliography{zs}

\end{document}